# Reflection and refraction in Robert Grosseteste's
# De Lineis, Angulis et Figuris

**Amelia Carolina Sparavigna**
Department of Applied Science and Technology, Politecnico di Torino, Italy

*In his scientific treatise entitled De Lineis, Angulis et Figuris, seu Fractionibus et Reflexionibus Radiorum, Robert Grosseteste is discussing some qualitative geometric rules about reflection and refraction. However, he is also discussing about the power of reflected and refracted light. The reading of these discussions is quite interesting, and what Grosseteste is telling seems well-posed, when compared with the Fresnel reflectance.*

Robert Grosseteste (c. 1175 – 1253), English scientist, philosopher, and bishop of Lincoln, is considered the origin of the revival in the study of optics during the XIII century in Europe. He is usually referred for his use of geometry to describe optics, for instance in the reflection and refraction of light. However, besides the geometry, A.C. Crombie in [1] is remarking that Grosseteste developed an analysis of the powers propagated from the natural agents. This analysis is found in four related essays written most probably in the period 1231-1235. Two of them are the De Colore [2] and the De Iride [3], that we have already discussed.
Another treatise is that entitled De Lineis, Angulis et Figuris seu Fractionibus et Reflexionibus Radiorum. Crombie shortly commented this treatise in such a manner: according to Grosseteste "the same power produced a physical effect in an inanimate body and a sensation in an animate one. He established rules for operation of powers: for example the power was greater for shorter and straighter the line, the smaller the incident angle, the shorter the three-dimensional pyramid or cone; every agent multiplied its power spherically. Grosseteste discussed the laws of reflection and refraction (evidently taken from Ptolemy) and their causes, and went on in De Natura Locorum to use Ptolemy's rules and construction with plane surfaces to explain refraction by a spherical burning glass" [1]. Let us remark however, that Grosseteste used the optics of Alhazen and Alkindi [4], besides that of Ptolemy.
What Crombie is telling about power of rays is stimulating to read the Grosseteste's treatise. Here I am translating and discussing the Latin text, subdividing it is several sections. The Latin text is in MS Reference Sans Serif characters, its translation in *italic*. For my translation, the Latin source is in Ref.[5]. We will see that the discussion on the power of reflected and refracted rays is rather complex, in any case, it seems well-posed when compared with the rigorous approach by Fresnel reflectance formulas.

Here the incipit and end of the treatise.

> INC: Utilitas considerationis linearum, angulorum et figurarum est maxima, quoniam impossibile est sciri naturalem philosophiam sine illis.

*Incipit: The utility of considering lines, angles and figures is huge, because it is impossible to know the philosophy of nature without them.*

> EXPL: ... potest dici, quod istae rationes optime concludunt, quantum sufficiunt, et ideo procederent, nisi rationes fortiores essent in contrarium, quae praedictae sunt. Explicit tractatus Lincolniensis de fractionibus et reflexionibus radiorum.

*End: ... it can be said that these reasons seems rather well posed, and they could be, if there were not the strongest reasons to the contrary, which we have mentioned previously. This is the end of the treatise by a Lincolnian on the reflections and refractions of rays.*

> 1 - Utilitas considerationis linearum, angulorum et figurarum est maxima, quoniam impossibile est sciri naturalem philosophiam sine illis. Valent autem in toto universo et partibus eius absolute. Valent etiam in proprietatibus relatis, sicut in motu recto et circulari. Valent quidem in actione et passione, et hoc sive sit in materiam sive in sensum; et hoc sive in sensum visus, secundum quod occurrit, sive in alios sensus in quorum actione oportet addere alia super ea, quae faciunt visum.

*The utility of considering lines, angles and figures is huge, because it is impossible to know the philosophy of Nature without them. They are valid for the entire universe and, unconditionally, for all its parts. They apply in connecting the properties, such as in straight and circular motions. And they apply in action and passion (reaction), and this is so, whether in the matter or in the capacities of perception; and this is so again, whether in the sense of sight, as it is occurring, or in any other sense in the action of which, it is necessary to add other things on that which is producing the vision.*

Let us report, after this strong incipit what A.G. Padgett is telling about Grosseteste in [6]. "Even as he translated and interpreted Aristotle, Grosseteste placed Aristotelian natural philosophy in a broader Christian and Neo-platonic world view. … he was committed to a natural philosophy based upon mathematics. This emphasis derived from Platonic and Pythagorean traditions, as mediated to him through Patristic authors like Augustine. A mathematical natural philosophy is demonstrated in a number of his works, particularly works on astronomy, light, and in his treatise on geometry, De Lineis, Angulis et Figuris." As we will see in reading this treatise, it is not a treatise on geometry, as told by Padgett, but on the geometry applied to light propagation. Padgett continues in [6] telling that in the incipit of the treatise, Grosseteste defends his mathematical approach to natural philosophy. "Notice that Grosseteste wants to use geometry, which was long a key tool of astronomers, within natural philosophy. This is a decisive step in the history of Western science, although Grosseteste was not alone in making it." [6].

> 2 - Cum igitur in aliis dictum est de eis quae pertinent ad totum universum et partes eius absolute, et de his quae ad motum rectum et circularem consequuntur, nunc dicendum est de actione universali, prout ipsa recipit naturam inferiorum; quae est subiectum susceptivum diversorum actuum, prout ad actionem in materiam mundi contingit descendere; possuntque aliqua in medium adduci, quae erudire possunt procedentem ad maiora.

*Then, since we have discussed elsewhere of those things pertaining to the whole universe and to its parts in an absolute sense, and of those which are consequent to straight and circular motions, now we have to tell something concerning the universal action, when it is receiving a lower nature; this universal action is a player able of various features, so far as it happens when it is descending to act in the matter of the world; moreover, other things can be questioned, that can educate us to proceed ad majora.*

In a translation by E. Grant [7], we find that Grosseteste is proposing a universal action descending in the lower world, according to an Aristotelian view of the universe. From [8], we find that "in medium rerum adducere," means to question, questioning. [8] "Ad majora" is a wish we can give to a person, to have greater things, that is, success and satisfaction.

> 3 - Omnes enim causae effectuum naturalium habent dari per lineas, angulos et figuras. Aliter enim impossibile est sciri "propter quid" in illis. Quod manifestum sic: Agens naturale multiplicat virtutem suam a se usque in patiens, sive agat in sensum, sive in materiam. Quae virtus aliquando vocatur species, aliquando similitudo, et idem est, quocunque modo vocetur; et idem immittet in sensum et idem in materiam, sive contrarium, ut calidum idem immittit in tactum et in frigidum.

*Therefore, all the causes of the natural effects must be given by lines, angles and figures, because it is impossible to know in another manner the "propter quid" in them. It is clear the following: a natural agent propagates (multiplies) its power from itself to the patient, the person or thing that undergoes some action, that is, whether it is acting on sense or on matter. This virtue is sometimes called "species", sometimes "likeness", and it is the same, in any way we call it; and the same thing is instilled in the sense and in the matter, or vice versa, when heat makes warm to the touch and gives itself to the cold body.*

*Species* in Latin means**:** *seeing, view, look; sight*; but also *external appearance; general outline or shape*. Then the species is that feature of the power of light which allows perceiving the shape of an object.

It the De Iride [3], we found the *quid* (interrogative pronoun [9]), *what*, that is the effect, or the phenomenon, the physics needs to describe, and the *propter quid, because of what*, which is instead an answer given by the research, on the causes of the phenomenon. And here Grosseteste is telling that without the geometry we are not able to answer.

In [7], *virtus*, *virtue*, is translated as *power*: in further reading the text, we will see that it is so. From Ref. [10], we find that "virtue" acquired in the early 13century the meaning of "moral life and conduct, moral excellence," from Latin *virtutem* (nom. *virtus*) "moral strength, manliness, valor, excellence, worth," from *vir* "man". However [10] is telling that phrase *by virtue of* (early 13c.) preserves alternative Middle English sense of "efficacy." For instance, the Wyclif Bible has *virtue* where King James Version uses *power*.

> 4 - Non enim agit per deliberationem et electionem; et ideo uno modo agit, quicquid occurrat, sive sit sensus, sive sit aliud, sive animatum, sive inanimatum. Sed propter diversitatem patientis diversificantur effectus. In sensu enim ista virtus recepta facit operationem spiritualem quodammodo et nobiliorem; in contrario, sive in materia, facit operationem materialem, sicut sol per eandem virtutem in diversis passis diversos producit effectus. Constringit enim lutum et dissolvit glaciem.

*For, it does not act through deliberation and choice; and therefore in one way it acts, whatever it is occurring, whether it is a perception or something else, animated or inanimate. But because of the diversity of the objects of action we have different effects. Moreover, in the perception, this received power produces, in some way, a spiritual and noble effect; on the other hand, when acting on the matter, it produces a material effect, such as the sun produces, through the same power, different effects in different objects of its action. For it hardens the clay and melts the ice.*

> 5 - Virtus igitur ab agente naturali aut veniet super lineam breviorem, et tunc magis est activa, quia patiens minus distat ab agente, aut super lineam longiorem, et tunc minus est activa, quia patiens magis distat. Sed sive sic, sive non sic: aut veniet immediate a superficie agentis, aut mediate. Si immediate: aut per lineam rectam veniet, aut per obliquam. Sed si per lineam rectam: tunc est actio fortior et melior, ut vult Aristoteles V Physicorum, quia natura operatur breviori modo, quo potest. Sed linea recta omnium est brevissima, ut ibidem dicit.

*Moreover, the power produced by a natural agent can come along a shorter line, and then, it is more active, because the patient receiving it is less distant from the agent, or it can move along a longer line, and then it is less active, because the patient is more distant. And the power can comes directly from the surface of an agent, or with mediation. If it comes without mediation, it can come by a straight line or by an oblique line. If, however, it comes by a straight line, then there is a stronger and better action, as Aristotle assumes in V Physics, because the Nature acts in the shorter available way. But the straight line is the shortest of all, as he says in the same book.*

> 6 - Item, linea recta habet aequalitatem sine angulo; sed melius est aequale, quam inaequale, ut dicit Boethius in arithmetica sua. Sed natura operatur breviori et meliori modo, quo potest; quare melius operatur super lineam rectam.

*Similarly, a straight line has equality and no angles; but equal is better than unequal, as Boethius [11] tells in his Arithmetic. And Nature acts in the possible shorter and better way, and therefore it works better on a straight line.*

Here we can repeat what Grosseteste is telling in De Iride [3]: Et idem manifestavit nobis hoc principium philosophiae naturalis, scilicet quod "omnis operatio naturae est modo finitissimo, ordinatissimo, brevissimo et optimo, quo ei possibile est". And the same tells us that principle of the philosophy of nature, namely, that "every action of the nature is well established, most ordinate, in the best and shortest manner, as it is possible."

> 7 - Item, omnis virtus unita est fortioris operationis. Sed maior unio et unitas est in linea recta quam in non recta, sicut dicitur in V Metaphysicae. Quare fortior erit operatio super lineam rectam.

*Again, every compact power is stronger in its operations. But, the greater union and unity is in a straight line rather than in distorted line, as stated in V Metaphysics. And then an action works stronger on a straight line.*

> 8 - Sed linea recta aut cadit ad angulos aequales, quae est perpendicularis, aut inaequales. Si ad angulos aequales, est operatio fortior propter tres rationes praedictas, quia illa linea est brevior et aequalis et virtus uniformiter venit per eam ad partes patientis.

*The straight line can fall either at equal angles, that is, it is perpendicular to the surface, or at unequal angles. If it falls at equal angles, the operation is stronger for the three abovementioned reasons, because the line is shorter and equal and its power comes more uniform through it to the parts of the patient, person or thing that undergoes the action.*

It seems that Grosseteste is talking about illumination. And here then, it is suitable to remember the cosine law of illumination, which is a geometric relationship between the illuminance of a surface and the angle of incidence of the illuminating rays. If a source of light is point-like, the illuminance that it produces on a surface depends on intensity, distance and angle of incidence. Then, let us consider the intensity I of the light in a particular direction from the source: the light travelling a distance d falling with an angle $\theta$, measured from the normal to the surface, has an illuminance E given by $E = I \cos\theta / d^2 = I \cos^3\theta / h^2$, where h is the perpendicular distance [12,13]. The maximum illuminance is for normal propagation.
Illuminance is analogous to irradiance, but is to be distinguished from the latter in that it refers only to light. A distinction is necessary between illuminance and luminance: the latter is a measure of the light coming from a surface.

> 9 - Linea autem cadens ad angulos aequales super corpus aliquod cadit ad rectos, quando cadit super planum; quando super concavum, ad acutos; quando autem super sphaeram, ad maiores recto. Quod manifestum est, quia, si ducatur linea incidentiae transiens per medium sphaerae cum linea contingentiae, facit angulos rectos, et ex linea contingentiae cum sphaera causantur utrobique anguli contingentiae; quare linea cadens super sphaeram faciet angulos duos cum eius superficie, quorum uterque est maior recto, quia valet angulum rectum et angulum contingentiae. Quando ergo cadit virtus ad angulos non solum aequales, sed omnino rectos, tunc videtur esse actio fortissima, quoniam omnino est aequalitas et uniformitas completa.

*A line, however, is falling down with equal angles on a body perpendicularly, that is with right angles, when it falls on a plane; when it falls on a concave body, it is at acute angles; but when it is over the sphere, it happens at angles larger than the right angle. This is shown as in the following, because, if a line is drawn passing through the center of a sphere, it makes a right angle with the line of contingency (tangency), and the line of tangency makes with the sphere on both sides the angles of contingency; then, the line falling on the sphere makes two angles with its surface, each angle larger than the right angle, being it the sum of the right angle and the angle of contingency. Thus when the power falls with angles which are not only equal, but right, then it would seem the action to be very strong, because there is complete equality and uniformity.*

Here it was quite helpful the discussion in [7], which tells that the Medieval scientists regarded "contingent angles", that is the angles of tangency, as having a finite magnitude. Therefore the contingent angle is different if it is of a convex or concave surface (nobody is perfect…).

> 10 - Si vero sit linea non recta, sive curva, tunc, cum non sit circularis, quia agens naturale, non facit virtutem suam secundum circulum, sed secundum diametrum circuli propter brevitatem, manifestum est, quod talis linea habebit angulos. Et hoc non fiet, dum medium est unum, sive dum est unum corpus occurrens; sed oportet, quod sint duo, unde in primo multiplicatur virtus super lineas rectas, in secundo super alias.

*If, however, it is not a straight line but it is a curve, nevertheless, not circular, because a natural agent does not produce its own strength according to a circle, but according to the diameter of the circle for the sake of brevity, it is manifest that such a line will have some angles. And this will not occur, as long as there is a single medium, or while there only one body; but it is necessary that two media exist, whence in the first the power is propagated along some straight lines, in the second along other lines.*

To bend the light we need several different media, so that at the interfaces the ray is broken with certain angles. This is discussed also in the De Iride [3], where we find even a law of refraction, which tells that the angles of refraction are one half the angles of incidence.
Grosseteste is telling that the power "multiplies" along a straight line. And in fact, when Grosseteste talks about the light and its propagation, he imagines it as multiplying itself [15]. I translated as he imagined the propagation of light as a multiplication, more or less, as proposed by Huygens for the waves. In 1678, Christiaan Huygens proposed that each point of a luminous wavefront can be the source of a spherical wavelet. The sum of these wavelets determines the new propagated wavefront. He assumed that the secondary waves travelled only in the forward direction. And then the light is "generating" itself, in the sense of propagation. May be, Grosseteste imagines a similar mechanics, without waves, however.

> 11 - Hoc autem non potest esse, nisi duobus modis: aut scilicet quod corpus patientis sit densum, ita ut impediat transitum virtutis, praecipue quantum ad sensum nostrum et tunc dicitur linea reflexa, ideo quod redit virtus aut corpus occurrens sit rarum, quod permittit transitum virtutis. Si primo modo, tunc virtus veniens ad corpus densum aut cadit ad angulos aequales sive perpendiculariter, aut ad inaequales. Si primo modo, tunc redit in se per eandem viam, per quam venit. Cuius ratio est, quia qualem angulum constituit linea cadens super corpus, talem et tantum constituit linea reflexa.

*This can happen only in two manners. First manner: that the body of the patient is dense, so as to impede the transit of power, especially in regard to our perception, and then it is said we have a reflected line, which is turning back the power. Second manner: the body the light is passing through is thin in density, which allows the propagation of power. If we have the first case, then we have the ray falling on a dense body, it falls with equal angles, that is, perpendicularly to the body, or with unequal angles, that is inclined. If we have the first manner, then it returns into itself through the same path, along which it arrived to the body. The reason of this is due to the*

*following, the line falling on the body makes such an angle, as it is the angle made by the reflected line.*

This is the law of reflection, telling the incidence and reflected angles are equal.

> 12 - Et ideo oportet, quod ad eosdem angulos reflectatur, super quos cecidit et per eandem viam redeat. Si enim rediret per alios angulos sive per aliam viam declinando a dextris sive a sinistris, impossibile esset, quod ad aequalem angulum cum angulo incidentiae rediret; faceret enim maiorem vel minorem.

*And therefore it is proper that it is reflected at the same angle, upon which the ray travelled and return by the same pattern, because if it were redirected with another angle or following another pattern, turning to the left or to the right, it would be impossible that the return forms an angle equal to the angle of incidence; it would be larger or smaller.*

> 13 - Si vero cadit ad angulos inaequales, tunc redibit per talem viam, qua possit facere angulum aequalem cum superficie corporis resistentis angulo incidentiae, scilicet illi, quem constituit linea incidens cum illo corpore propter rationem praedictam. Universaliter enim angulus incidentiae et reflexionis facit angulos aequales, quod supponatur nunc.

*In the case that the ray is not falling perpendicularly, then it comes back along such pattern, able to make an angle with the surface of the resisting body equal to the angle of incidence, namely, the angle which is made by the incident line with that body, for the argument already mentioned. Generally speaking, the angle of incidence and the angle of reflection are equal, which is to be assumed now.*

> 14 -Cum ergo his duobus modis fiat reflexio, intelligendum est, quod virtus reflexa in se propter geminationem virtutis in eodem loco fortior est, quam virtus reflexa in aliam viam. Attamen, quantum est de ratione reflexionis, debilior est actio, ubi est reflexio in eadem via, quia, cum omnis reflexio debilitet virtutem, illa tamen, quae facit declinare omnino virtutem ab incessu recto, quem deberet habere, si per medium corporis transiret, magis debilitatur; et haec est, quae est in eadem via, a qua venit. Haec enim via est omnino contraria et opposita incessu recto, quem deberet habere.

*Since these are the two modes in which reflection can happens, it is to be understood that the reflected power into itself, because of a doubling of the power in the same place, is stronger than the reflected power in another path. Nevertheless, and this is in the essence of reflection, the action of the reflected ray is weaker, when there is the reflection in the same path, since each reflection is weakening the power, and this precise reflection, making the power to have a complete deviation of 180° from the straight prolongation of the incident ray (that is, the direction the ray would have if it were to pass through the body), is highly weakened; and this is for the ray, which is in on the same path on which it came from. Moreover, the path is totally contrary and opposed to the incident one, as it is to be.*

First of all, let us discuss the first sentence, that on the "genimatio", doubling. A possible interpretation can be the following: let us consider a ray of light normally incident on a surface and the reflected ray, radiated back into the half-space of the incident ray. It means that in the volume occupied by these rays, which is the same, we have a "doubling", a superposition of power. In any other case, that is, when the incidence is oblique, a certain volume of the space can be occupied just by the incident or by the reflected ray.

Grosseteste continues discussing the power of the reflected rays as depending on the angle of incidence. Here is quite useful the suggestion of a deviation of 180° given in Ref.7. What is told by Grosseteste is in agreement with the fact that the light falling at an angle on a surface tends to be increasingly reflected as the angle of incidence increases, and the transmission reduced. For a normal incidence, in fact, we have the largest amount of transmitted power and the smallest reflected. Usually, the behavior of the reflected light with the angle of incidence is studied according to polarity. The Maxwell's equations allows the derivation of the Fresnel equations (see for instance, the Fresnel laws of reflection as discussed by the Chapter 33, Volume 1 of the Feynman Lectures on Physics), which can be used to predict how much of the light is reflected and refracted. On a specular reflection then, we have that the fraction of the reflected light increases with increasing angle of incidence θ.

The Fresnel reflectance for metals and dielectric materials is very different. For a metal such as aluminum, the reflectance is always above the 85%. For a glass having a refractive index of n=1.5, the reflectance is of only 4% at normal incidence, but 100% at grazing. "This effect, in fact, is what makes polished metals look like metal, and polished glass not look that way. It's also why it's hard to comb your hair in a shop window; you are looking at the angle of minimum reflectance." [14]

> 15 - Quod si fiat reflexio a corporibus politis habentibus naturam speculi, tunc est optima reflexio et fortior actio; cum vero a corporibus asperis, tunc dissipatur species et actio est debilis. Cuius causam assignat Commentator super tractatum de sono dicens, quod partes corporis politi et levis superficiei propter suam aequalitatem et uniformitatem concurrunt omnes in unam actionem in reflexione speciei; et ideo tota integra, sicut venit, reflectitur a corpore polito. Sed partes corporis asperi sunt inaequales et quae altior est, primo reflectit speciem; et ideo non concordant partes in unam actionem, et propter hoc dissipatur species in partes, et ideo non est bonae operationis.

*When we have a reflection from some bodies polished to have the same nature of mirrors, then we have the best reflection and stronger action; but when reflection happens on rough bodies, the species, that is, what is making the appearance of an object to the sight, are dissipated, and the action is weak. The reason is given by Averroes, the Aristotle's Commentator, in his discussion on the sound, saying that the parts of a body surface smooth and polished, for its equality and uniformity, all together are concurring into a single action in the reflection of the species; and therefore the whole power, as it came, is reflected back from the polished body. But when the parts of a rough body are unequal,  those parts protruding are reflecting the species first, and therefore there is not an agreement of the parts in a unique action, and for this reason we have a dispersion of this species randomly, and this is not a good operation.*

Here Grosseteste is distinguishing between specular and Lambertian surfaces. Very interesting is the fact that Grosseteste is using an analogy with the sound waves, telling that Averroe studied the sound propagation and the role of irregular surfaces in break down the reflection of it.

> 16 - Quando etiam est reflexio a corporibus concavis, maior est actio, quam a planis et convexis, eo quod radii reflexi a superficie concava concurrunt in unum; non autem sic de aliis.

*When the reflection is obtained by means of some concave bodies, the action is stronger, than when the bodies are plane or convex, and this happens because the rays reflected by a concave surface converge together; this does not happens for the other cases.*

> 17 - Si vero corpus concurrens non impediat transitum virtutis, tunc radius cadens ad angulos aequales sive perpendiculariter tenet incessum rectum et est fortissimus. Sed ille, qui cadit ad angulos inaequales, deviat ab incessu recto, quem habuit in corpore priore et quem deberet adhuc habere, si esset medium uniforme. Et ista deviatio vocatur fractio radii.

*Indeed, if the medium encountered by the light is not impeding the transit of power, a ray incident at equal angles, that is perpendicularly, maintains the straight line and is the strongest ray. But the ray, which is incident at unequal angles, that is inclined, deviates from the straight line that the ray*

*had in the first medium and that it would still have if the medium were homogenous. This deviation is called refraction of rays.*

For a normal incidence we have the largest amount of transmitted power. The transmitted power is reduced increasing the incidence angle. See the discussion of the section 14.

> 18 - Et haec est dupliciter: Quoniam si illud corpus secundum est densius primo, tunc radius frangitur ad dexteram et vadit inter incessum rectum et perpendicularem ducendam a loco fractionis super illud corpus secundum. Si vero sit corpus subtilius, tunc frangitur versus sinistrum recedendo a perpendiculari ultra incessum rectum. Et cum haec sint ita, intelligendum est tunc, quod virtus veniens super lineam fractam fortior est, quam super reflexam, quia linea fracta parum recedit ab incessu recto, qui est fortissimus, et reflexa linea multum recedit in oppositam partem, unde reflexio plus debilitat virtutem quam fractio.

*The refraction is twofold: when the second medium is denser than the first, the ray is refracted to the right and passes between the prolongation of the direction of incidence and the perpendicular drawn from the point of incidence in the second medium. When the second medium is rarer, the ray is refracted to the left, receding from the perpendicular beyond the prolongation of the incident ray. And then, since these are the facts, we need to understand the reason why the power incident along a refracted line is higher that the power along a reflected ray, this happens because a refracted line little deviates from the prolongation of the incident ray, which is the strongest, and the reflected line largely deviates in the opposite direction, and then the reflection is weakening the power more than refraction.*

In the first part of this section we have a sort of Snell law of refraction. For what concerns the power, to have an agreement of the last sentence with what was previously told on the intensity of transmitted and reflected light, we have to assume that Grosseteste is considering a normal incidence or an incidence at small angles.

> 19 - De virtute autem fracta dupliciter potest dici, quod virtus fracta a dextris fortior est quam ista a sinistris, eo quod ista, quae frangitur a dextris, magis accedit ad incessum perpendicularem, sive loquamur de illa perpendiculari, quae ducitur a loco fractionis, sive ab agente, a cuius puncto eodem exeunt linea perpendicularis et linea fracta.

*About the power of the two modes of refraction we can tell that the power refracted to right is greater than that refracted to left, since this power, that to the right, is closer the perpendicular to the interface, whether this is the perpendicular line drawn from the incidence point, or a line drawn from the agent, from which the perpendicular line and the refracted line have their origin.*

May be, he was arguing on the total internal reflection.

> 20 - Praeter vero istas tres lineas essentiales est quarta accidentalis, super quam venit virtus accidentalis et debilis. Quae quidem venit non ab agente immediate, sed a virtute multiplicata secundum aliquam trium linearum dictarum; secundum quod a radio cadenti per fenestras venit lumen accidentale ad omnes angulos domus. Ista autem virtus est omnium debilissima, quoniam non ab agente exit immediate, sed a virtute agentis fracta secundum lineam rectam, reflexam vel fractam. Haec igitur de lineis et angulis dicantur.

*Besides these three fundamental lines, there is a fourth accidental line, along which an accidental and weak power moves. Which, indeed, does not come directly from an agent, but is coming from a power propagated by any of the three abovementioned lines; in such a manner, from a ray entering a window, it comes, by chance, the light to all the corners of a house. However, this power is the*

*weakest one, because it does not come directly from the agent, but it is separated from the power of the agent, in a straight line, or reflected or refracted. These facts we told about lines and angles.*

> 21 -De figuris autem duae species ad praesens considerandae sunt. Quarum una est necessaria propter multiplicationem virtutis, scilicet sphaerica. Omne enim agens multiplicat suam virtutem sphaerice, quoniam undique et in omnes diametros: sursum deorsum, ante retro, dextrorsum sinistrorsum. Quod patet per hoc, quod, qua ratione ab agente posito loco centri contingit protrahere lineam in unam partem, et in omnem secundum omnes differentias positionis; quapropter oportet, quod sphaerice. Ita dicit Commentator super secundum de anima. Item ubicunque ponatur sensus, potest sentire tale agens in distantia debita; sed non nisi per speciem sive virtutem venientem ab agente. Illa ergo virtus undique multiplicatur.

*About the figures, there are two kinds of them that we have to consider here. One of these is suitable for propagation of power, namely the sphere. And this for the following reason: every agent emanates its power spherically, since it does all around and in every direction (diameter): upwards and downwards, ahead and aback, right and left. And this is shown by the manner in which it is possible to draw a line in a certain direction from an agent located at the center, and in all directions from all the different positions, and therefore it is proper to use that spherical figure. And this is in agreement with what the Commentator (Averroes) says in the (Aristotle's) De Anima. Also, wherever we put the sensor to receive, we can feel such an agent at a proper distance; however this happens only by species or by the power coming from the agent. So the power is propagating everywhere.*

Here again, instead of "multiplication", I prefer "propagation".

> 22 - Alia autem figura exigitur ad actionem naturalem, scilicet pyramidalis: quoniam si virtus veniat ab una parte agentis et terminetur ad aliam partem patientis et sic de omnibus, ita quod semper veniat virtus ab una parte agentis ad unam solam partem patientis, nunquam erit fortis actio sive bona. Sed completa est actio, quando ab omnibus punctis agentis sive a tota superficie eius veniet virtus agentis ad quemlibet punctum patientis. Hoc autem est impossibile, nisi sub figura pyramidali, quoniam virtutes venientes a singulis partibus agentis concurrunt in cono pyramidis et congregantur et ideo omnes fortiter possunt agere in partem patientis concurrentem.

*Another figure, however, is required for the natural action, that is, the pyramidal one: since, if the power is coming out from a part of the agent and ending onto another part of the patient, and so on for all the parts of agent and patient, we always had the power from a part of the agent falling onto a sole part of the patient, the action will never be strong or good. However, the action is complete when the power of the agent comes from all the points of the agent or from its whole surface to every point of the patient. But this is impossible, except under the pyramidal figure, because the powers coming from each part of the agent are concurring in the cone of the pyramids and gathered together and then they all are able to act more strongly upon the part of the patient where they are condensed.*

That is, instead of a disordered analysis of the propagation of some rays, it is better to consider the solid angles.

> 23 - Possunt ergo infinitae pyramides exire ab una superficie agentis, quarum omnium una est basis, scilicet superficies agentis, et coni sunt tot, quot sunt pyramides, et cadunt in diversa puncta medii seu patientis undique; et ad unam partem possunt infinitae exire, quarum una est brevior et alia longior. Sed illae, quae sunt aequalis longitudinis et brevitatis, non habent diversitatem, quia aequaliter agunt, quantum est ex parte sua, licet varietas possit esse a parte materiae recipientis.

*Therefore, an infinite number of pyramids can come out from a surface of an agent, pyramids having the same basis, namely, the surface of the agent, and there are so many cones as the pyramids are, falling into different points of the medium or on all sides of the patient, and there can be an infinite number coming out from surface, some shorter some longer. However, those cones which are equal in length and size, do not possess different features, because they act in the same manner, though there can be a variety of features coming from the recipient matter, inasmuch it is concerning it.*

> 24 - Quando autem una est brevior alia, et exeunt ab eodem agente, pulchra est difficultas, utrum conus pyramidis brevioris magis agat in patiens? Et oportet ponere, quod pyramis brevior magis agat, quia conus eius minus distat a fonte suo, et ideo plus virtutis ibi invenitur, quam in pyramide longiori et ideo patiens a pyramide breviori est magis coniunctus agenti et ideo fortius alteratur secundum virtutem.

*But when one pyramid is shorter than the other, and both are coming out by the same agent, we have a quite difficult problem to solve, that of telling whether is the cone of the shorter pyramid acting more on the patient or not. And then, we ought to suppose that the shorter pyramid acts more, because its cone is less distant from its source, and for that reason, there is more power in it than in the longer pyramid and then the patient is more closely connected to the agent and therefore strongly altered by its power.*

In Ref.[4], it is told that in the De multiplicatione Bacon says that shorter pyramids act with more strength than longer pyramids, because the vertex is less distant from the agent. Therefore "the virtue coming by means of it is less weakened." According to [4], Grosseteste was then probably Bacon's source of his idea.

> 25 - Praeterea si radii, qui sunt de corpore pyramidis brevioris, qui veniunt a dextra parte, protrahantur ultra conum in continuum et directum, facient minores angulos cum radiis sinistris, qui sunt de corpore pyramidis, quam radii similes, qui sunt a parte pyramidis longioris, ut patet per 21um, primi geometriae Euclidis et etiam ad sensum. Et eodem modo est de radiis venientibus a sinistra parte pyramidis, exeuntibus ultra conum in continuum et directum, qui magis coniunguntur cum radiis dextris, qui sunt de corpore pyramidis, quam consimiles faciunt a parte pyramidis longioris.

*Moreover, if the rays which are in the bulk of a shorter pyramid, that come from the right side, are prolonged besides the vertex, uninterrupted and straight, they will form smaller angles with the left beams, which are in the bulk of the pyramid, than the similar rays which are coming from a longer pyramid, as it is clear from the 21th section of first book of Euclid Geometry, and also by the common sense. And in the same way, the rays coming from the left of the pyramid, which continues beyond the vertex, uninterrupted and straight, are closer to the rays of the right side in the bulk of the pyramid, than the consimilar rays of a longer pyramids.*

> 26 - Quapropter, cum omnis congregatio et unitio magis est activa, conus pyramidis brevioris magis aget et etiam alterabit patiens quam longioris. Si tamen obiciatur rationabiliter, quod, cum a tota superficie agentis venit virtus ad conum pyramidis longioris, ubi virtus magis congregatur, propter hoc quod conus ille magis est acutus quam in breviori, et omnis virtus unita est maioris operationis, atque addatur ad haec, quod radii pyramidis longioris magis sunt propinqui radiis perpendicularibus ductis ab extremitatibus diametri agentis, quare sunt fortiores, quia incessus perpendicularis est fortissimus: potest dici, quod istae rationes optime concludunt, quantum sufficiunt, et ideo procederent, nisi rationes fortiores essent in contrarium, quae praedictae sunt. Explicit tractatus Lincolniensis de fractionibus et reflexionibus radiorum.

*Then, because any congregation or union is more active, the cone of a short pyramid acts more and alters the patient more than a longer cone. However, we could object rationally that, when from all the surface of an agent the power is coming in a longer pyramid, we have there more power, because the cone is more acute than that of a shorter pyramid, and all the power is condensed for a greater operation, and there is also to add the following, that the rays of a longer pyramid are close to the rays of the agent, those lines which are drawn perpendicularly from the ends of the diameters of the agent, and then they are stronger, because the perpendicular progression is the strongest: it can be said that these reasons seems rather well posed, and they could be, if there were not the strongest reasons to the contrary, which we have mentioned previously. This is the end of the treatise by a Lincolnian on the reflections and refractions of rays.*

In my opinion, Grosseteste used the solid angles to analyze emitted and received power.

Let us conclude with what is told in Ref.16 about the treatise we have here discussed. It is written there that the claim that Grosseteste gave a "special importance to mathematics in attempting to provide scientific explanations of the physical world is on a stronger footing", as we can find in the opening of On Lines, Angles and Figures. In the treatise, On the Nature of Places, which is its continuation, Grosseteste sums up the preceding text with the remark that "the diligent investigator of natural phenomena can give the causes of all natural effects, therefore, in this way by the rules and roots and foundations given from the power of geometry" [16,17]. Ref.16 continues telling that at the basis of the reasoning on light, there was Grosseteste's view that natural agents act by the multiplication of their power or species, a view developed further on by Roger Bacon. However, if we consider that the "multiplication" can be seen as propagation, this could be a sort of propagation as Huygens imagined several years after.
"Grosseteste holds that the intensity of operation of the natural agent will be a matter of its distance from what it acts upon, the angle at which it strikes it, and the figure in which it multiplies its operation, this being either a sphere or cone. He establishes certain rudimentary rules to this effect, such as that the shorter the distance, the stronger the operation": this is told by Reference 16. As we have seen from reading the Latin text, some observations on the power of transmitted and reflected light are more that rudimental. It seems that he had found these facts in some texts, probably from the Arab commentators of Aristotle, or, even, that he had experimented about them. May be, in a further reading on Grosseteste's treatises we can find an answer.